\documentclass{svjour3}                     % onecolumn (standard format)
\smartqed  % flush right qed marks, e.g. at end of proof

\usepackage{graphicx}
\usepackage{url}
\usepackage{algorithm2e}

\journalname{Social Network Analysis and Mining}
\begin{document}

\title{Communities and Hierarchical Structures in Dynamic Social Networks: Analysis and Visualization}


\author{Fr\'ed\'eric Gilbert   \and
				Paolo Simonetto \and
				Faraz Zaidi \and
				Fabien Jourdan \and
				Romain Bourqui
}			


\institute{F. Gilbert \and  P. Simonetto \and F. Zaidi \and R. Bourqui \at
              CNRS UMR5800 LaBRI and INRIA Bordeaux - Sud Ouest, France, \\
              Tel.: +33 540 00 84 27\\
              Fax:  +33 540 00 66 69 \\
              \email{\{frederic.gilbert, paolo.simonetto, faraz.zaidi, romain.bourqui\}@labri.fr}           %  \\
           \and
           F. Jourdan \at
              INRA, UMR1089, Toulouse, France \\
              fabien.jourdan@toulouse.inra.fr \\
              Tel.: +33 561 28 57 15               
}
\date{Received: date / Accepted: date}

\maketitle

\begin{abstract}
Detection of community structures in social networks has attracted lots of attention in the domain of sociology and behavioral sciences. Social networks also exhibit dynamic nature as these networks change continuously with the passage of time. Social networks might also present a hierarchical structure led by individuals that play important roles in a society such as Managers and Decision Makers. Detection and Visualization of these networks changing over time is a challenging problem where communities change as a function of events taking place in the society and the role people play in it.

In this paper we address these issues by presenting a system to analyze dynamic social networks. The proposed system is based on dynamic graph discretization and graph clustering. The system allows detection of major structural changes taking place in social communities over time and reveals hierarchies by identifying influential people in a social networks. We use two different data sets for the empirical evaluation and observe that our system helps to discover interesting facts about the social and hierarchical structures present in these social networks.

\end{abstract}

\section{Introduction}\label{intro}

A social network is a set of people connected by a set of social relationships \cite{wasserman:95,scott00} such as friendship \cite{rapoport61} and business collaboration \cite{watts98,newman01}. Mathematically these networks can be represented by a graph where \textit{nodes} represent people and \textit{edges} represent their relationships. Past work in social network analysis \cite{wasserman:95} has shown that the knowledge of community structure and relationship strength has important applications in web analytics \cite{chakrabarti:98}, marketing studies \cite{domingos:01}, homeland security \cite{yang07,subr08} and disease modeling \cite{kretzschmar96,eubank04}. 

Visual analysis of social networks is an integral component of the field of social network analysis \cite{freeman04}. Visualizing community structures present in social networks and identifying people who play important roles within a network can reveal interesting information specially by exploiting the temporal evolution of relationships. Social networks can exhibit temporal dynamics in a number of ways. The instances in the data may appear and disappear over time whereby different time windows may exhibit different characteristics. For example, a person might change his affiliation with a business organization by joining a different business enterprise and developing new social ties within this new environment. Moreover, the relationships may represent events and associations that are significant at a particular point of time, such as new job opportunities, or the establishment of a new business organization. If this is the case, then the temporal dimension associated with these events play a key role to capture important information. 

A more recent application of social network analysis has been in the study of counter terrorism \cite{memon07,maeno09,adler07,yang07}. Studying social networks of potential terrorists can help us to uncover the organizational structure of terrorist networks, predict terrorist activities by identifying events and possibly disclose the identity of master minds behind the criminal activities. 

This was the initial problem that motivated this research where we were required to analyze the data of cell phone calls (see section \ref{data} for more details). The goal was to analyze the dynamics taking place in social network over time and infer an influence hierarchy. The social network was represented by cell phone data where two people were connected if they communicated with each other through a cell phone . The initial work of this research was focused on this particular problem whereas we present an extended system in this paper which is generic and robust to handle a variety of data sets.

Other examples of dynamic social networks include email network \cite{diesner05}, where the time of an email sent, the co-authorship network of scientific publications \cite{newman01} with the year of publication and the actor-actor collaboration network of movies \cite{barabasi99} with its year of release. All these examples of social networks have temporal dimensions associated with them and must be exploited to analyze and understand these networks.

%What is in this Framework
In this paper, we present a system, called DySNAV abbreviate for \textbf{Dy}namic \textbf{S}ocial \textbf{N}etwork \textbf{A}nalysis and \textbf{V}isualization which helps a user to analyze the dynamics of community structures present in these social networks. People form community structures by frequently communicating or collaborating with certain people as compared to others. These communities undergo changes with the passage of time as the individuals, their relationships and their roles change in the social network. We try to identify these dynamics by focusing on communities and their changing relationships through visualization and discover important events by observing any radical changes in the structure of social network. We also infer a role hierarchy by identifying the most influential people in the social network. 

The paper is organized as follows: In the following section, we present the related work. In section~\ref{data}, we present different data sets used for experimentation. Section~\ref{proposed} presents the proposed system comprising of four major steps. The first step is data discretization described in section \ref{Discretization}. This is followed by the decomposition step in section \ref{decompo} where the community structures are identified. The details of how changes are detected in the community structures through visualization are presented in section \ref{detection}. In section~\ref{hierarchy}, we introduce a novel heuristic to determine the influence hierarchy in the network. As a case study, we use our system to analyze two dynamic social networks in section \ref{case}. Finally in section~\ref{conclusion}, we present conclusions and directions for future research.

\section{Related work}

Community detection in social networks has attracted lots of attention in the domain of sociology. A more generic formalism for the term \textit{community} is the term \textit{cluster}. Sociologists use the term \textit{community} \cite{coleman64} as compared to the statistical and data mining domain where people use the term \textit{cluster} \cite{tryon39} to refer to the same concept. A cluster might not necessarily represent a community but throughout this paper, we use the terms interchangeably to refer to the same concept. Several surveys \cite{jain99,berkhin02,schaeffer07} are available addressing the clustering or community detection problem. Some approaches \cite{ACJM03,newman04,girvan02} have performed better than the others for the discovery of communities in social networks. Researchers have also shown interest in discovering changing clusters in dynamic data \cite{Kalnis:05} and clustering evolving data streams \cite{aggarwal:03}. However, these techniques are either insufficient or inefficient to characterize the changes in community structures. Since the interactions taking place between individuals can be characterized by a single relationship (for example: a weighted edge), interactions between communities inherit a number of ways that can establish an interaction between two communities over a passage of time. Since most of the existing techniques are adapted to handle changes occurring in individuals rather than communities, the goal of our approach is clearly different from others. 
 
Social network visualization has also attracted much interest as images of social networks have provided investigators with new
insights about these networks\cite{freeman00}. Different visualization softwares and tools exist for social network analysis such as \cite{heer05,baur02,bilgic06,shen06} but these networks do not handle the temporal dimension of a network. The readers are recommended \cite{freeman00} for a more detailed review of the literature on social network visualization.

Research in the domain of analysis and visualization of dynamic graphs has attracted limited interest. For example, Kang \emph{et al.} \cite{kang07} introduce a tool called C-Group for temporal analysis of social networks. The tool focuses on a pair of individuals rather than analyzing overall structural changes in the entire network. Gloor \emph{et al.} \cite{gloor04} proposes a sliding time frame algorithm to display active ties between actors in a sliding time frame covering a time interval. The approach works well to trace the evolution of relationships between individuals but does not capture the evolution of the community structures in the entire social network. Sarkar and Moore \cite{sarkar05} present a method for modeling relationships that change over time. The idea is to develop an understanding of historical data and to predict future interactions. The model can be used to study the behavior of individual relationships but requires adaptation to model the behavior of a group of people. SoNIA (Social Network Image Animator)~\cite{demoll06} is a package for animating network dynamics over time and is not intended to be a network analysis tool. Rather than focusing on calculating network properties and indices, it is designed to facilitate the exploration of dynamic relational data, and the comparison of various layout techniques for making reliable animations of networks. It does not capture the dynamics of a group of people(cluster) and focuses on aggregating and transforming dynamic data to create a stable social space which is necessary to create a meaningful visualization. Moody \emph{et al.} \cite{moody05} introduce two types of visualizations: \textit{flip books} where nodes remain in a constant position and arcs fill in the holes among these nodes and \textit{dynamic movies} where nodes move as a function of relational changes taking place in the network. 

These systems perform well to exploit the temporal dimensions of a dynamic network focusing on changes and transitivity of individuals or their relationships. The system we present in this paper helps to discover structural changes in the entire network by studying the evolution of communities and the goals are clearly different from the other systems presented in this section.

\begin{figure*}[ht]
	\centering
	\includegraphics[width= .9\linewidth]{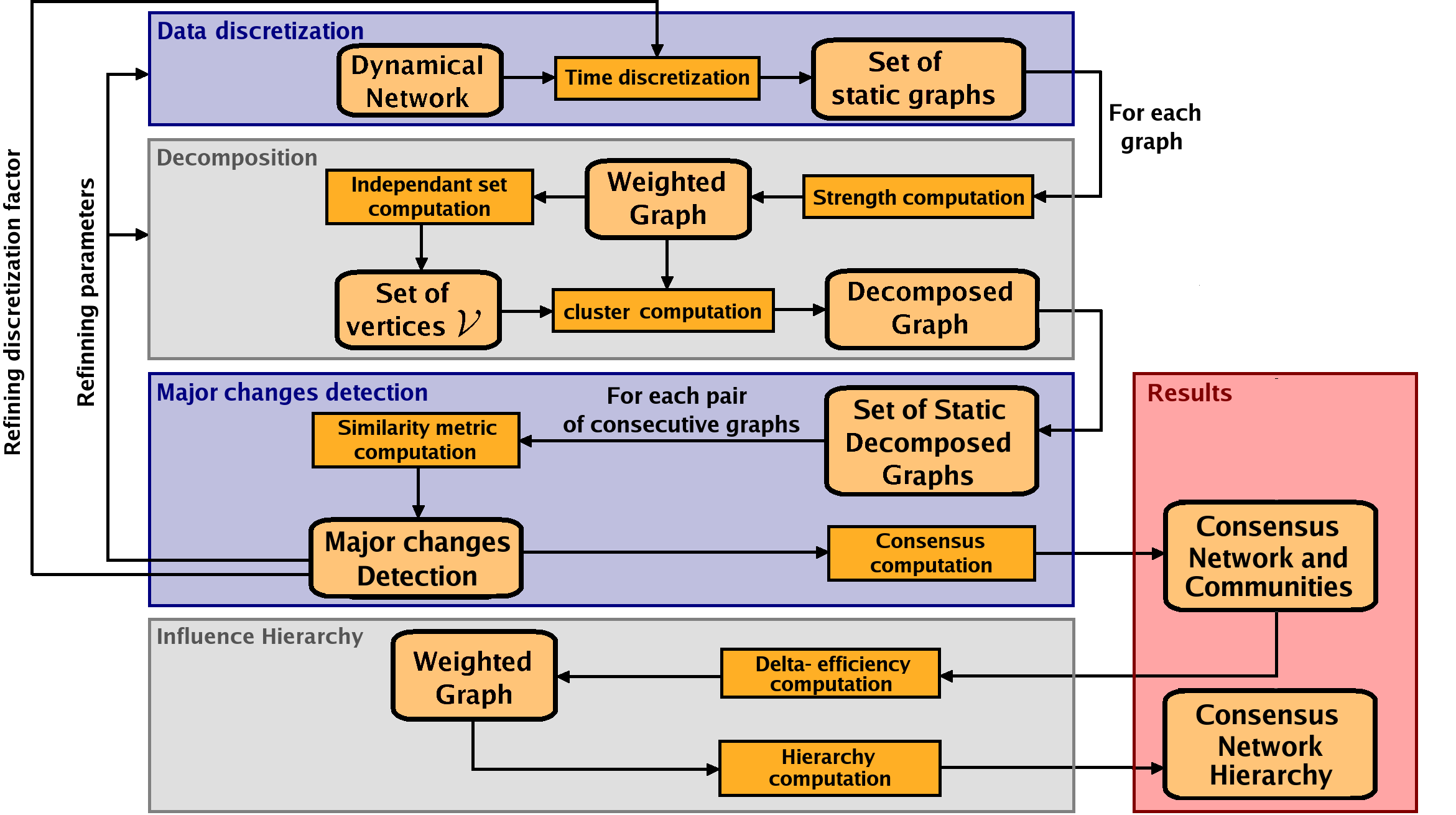}
	\caption{\label{fig_overview}Framework of the Proposed System representing the four major steps.}
\end{figure*}

\section{Data Sets}\label{data}

We use two different data sets for the empirical evaluation of our system. 

\textit{Catalano/Vidro} data set is a fictitious data presented in the IEEE VAST 2008 CHALLENGE \footnote{\url{http://www.cs.umd.edu/hcil/VASTchallenge08/}} for visualization and extraction of information about a terrorist group in the entire social network. It consists the information of $9834$ phone calls between $400$ cell phones over a $10$ day period in June 2006 in the Isla Del Sueno. The data set records each call as 5-tuple (from$\_$user$\_$id, to$\_$user$\_$id, timestamp, call$\_$duration, cell$\_$tower$\_$location). This is an interesting example as precise information about call records can be made available through any cell phone network. Tracking cell phone records with the associated temporal dimension can help us find or predict an event by an unexpected rise in the call frequency, distribution of important information, identity of people responsible for communicating information in the network etc.

The other example is the \textit{Co-Authorship Network} which is a network of researchers where two people are connected to each other if they have co-authored a scientific artifact. The year of publication is the temporal information associated to each artifact. The bibliographic data was downloaded from the DBLP Computer Science Bibliography website \footnote{\url{http://www.informatik.uni-trier.de/~ley/db/}} and contains data till the year 2008. From the complete data set, a subset was generated by selecting a researcher named \textit{Ulrik Brandes} and taking all the researchers connected to him at distance two i.e. the people who have directly co-authored with him, or have co-authored with a person having directly co-authored with \textit{Ulrik Brandes}. The data set is represented by 5-tuple (Author1,Author2,Year,Strength,Title\_of\_Artifact). The strength parameter was set to a default value of 1 for all entries. A complicated metric can be used such as if an artifact is co-authored by exactly two people, it will have a high strength whereas a high number of co-authors can represent a weak relationship between any two of its authors. The data set contains all the publications of \textit{Ulrik Brandes} available on the DBLP website from the year 1997 till the year 2008 containing approximately 900 researchers and 6500 edges between them.

\section{Proposed System}\label{proposed}

\begin{figure*}[ht]
	\centering
	\includegraphics[width= .99\linewidth]{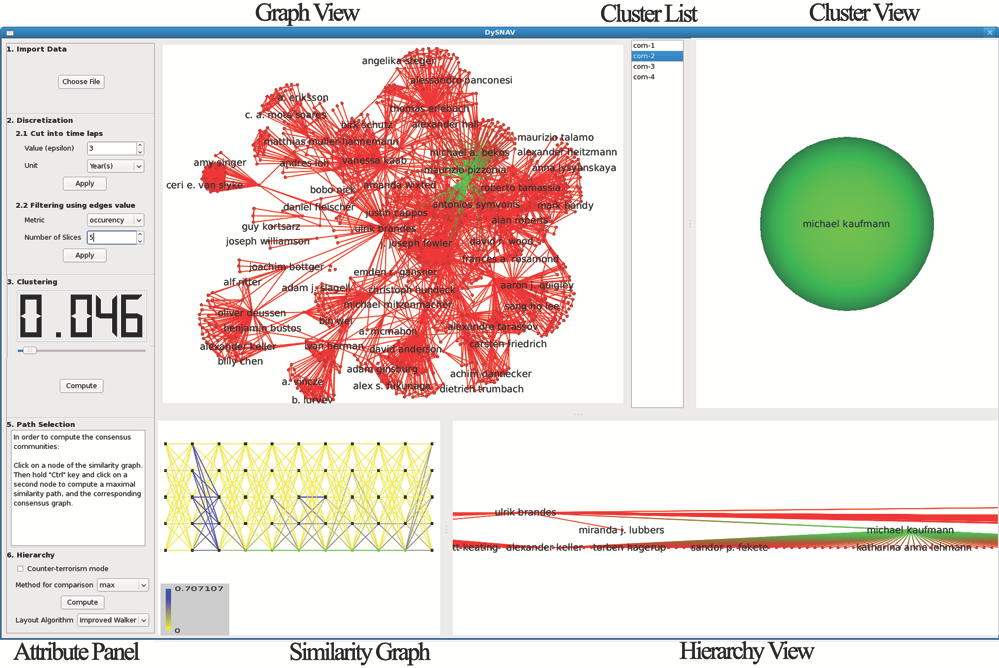}
	\caption{Screen Shot of the Proposed System. Different windows showing the various visualizations available in the system. Top Left: Graph for a selected interval with edges present in that time interval only. Bottom Left: Similarity Graph of different time intervals with time on the x-axis and different values of filter on the y-axis. Top Middle:List of all the clusters found in the Graph displayed in Graph View. Top Right: Contents of a Cluster. Bottom Right: Influence Hierarchy representing the most influential people closer to the root of the tree.}
	\label{fig_system}
\end{figure*}

A dynamic social network can be defined as a dynamic graph $G=(V,E)$ where $V$ represents the set of nodes (people) and $E$ represents the set of edges (relationships). Every edge $e=(u,v)\in E$ has an attribute depicting the Time described over a time period $[0 \dots T]$. The graph $G_{[t1,t2]}$ represents the nodes of the graph with only the edges present during the time interval $[t1,t2]: 0 \leq t1 < t2 \leq T$.

The main idea of the system is based on the framework introduced in \cite{bourqui09} proposed by the authors and was a preliminary version of this on going research. Figure~\ref{fig_overview} illustrates the four steps of the framework upon which this system is built. In this paper, we present a fully operational and interactive system based on the principles introduced in \cite{bourqui09} which is composed of several steps described below. 

The first step is to convert the dynamic graph into a set of static graphs where each static graph corresponds to a time interval. The discretization factor is taken as input which can be adjusted by the user interactively. 

The second step clusters each static graph separately using an overlapping clustering algorithm, to produce Fuzzy Clusters. This step allows us to identify communities in the network but also its \emph{pivots} (vertices shared by several clusters) while being insensitive to minor changes in the network as proven by \cite{bourqui09a}. 

The third step detects major structural changes in the network. We compare the clusterings obtained on every pair of successive static graphs using a similarity measure described in section~\ref{detection}. A low similarity indicates major changes during the period corresponding to the pair of snapshots while a high similarity value correspond to stable periods where the topological structure of the network does not go any major changes. Thus, once we have the similarity matrix from clusterings computed in step $2$, we can decompose the temporal changes in the input network into periods of high activity and \emph{consensus} communities during stable periods.

The last step consists of finding a \emph{role} or \emph{influence hierarchy} in the \emph{consensus} communities filtered from step $3$. We define the \emph{influence hierarchy} as a tree where the height of a node represents the influence of that node in the network. Our technique is based on the Delta efficiency metric \cite{ML06,larsen07} which computes the importance of a vertex with respect to the flow of information in the entire network. Using this Delta efficiency metric and Kruskal's algorithm~\cite{K56} for minimum spanning tree generation we are able to infer the influence hierarchy in the network.

The input to the system is a comma separated text file. Each line in the file is a 5-tuple (user\_id\_1, user\_id\_2, timestamp, relationship$\_$strength, relationship$\_$class). The user\_id\_1 and user\_id\_2 are identification numbers or strings used to identify two people in the social network. The timestamp is in the format yyyy/mm/dd-hr:mn:sc. The user is not required to enter all the information. For example if there is only year and months data available for a data set, the user can only enter the data in the format yyyy/mm in each tuple. In case if the data is available for a 24 hour interval with a precision of minutes, in that case, the user is required to enter all the information by entering the same year, month, and day data for all the entries like 2009/12/01-11:24 and change only the hour and minute information for other entries. The parameter relationship$\_$strength represents an integer value to assign a numerical weight that can be used as metric to distinguish between strong and weak relationships. As an example, for an email network, the size of the message can be used as an attribute of the relationship. Any default value can be assigned to all relationships if no real value exists. The parameter relationship$\_$class represents a nominal value to help classify relationships. Again considering the example of the email network, an IP address can be used as a class and any default value can be used for all the records. To load a data set in the system, the user is required to choose a file in the mentioned file format by clicking on the \textit{choose file} option in the \textit{attribute panel} of the system as can be seen in Figure~\ref{fig_system}.

The system comprises of five windows to display information and a panel (\textit{attribute panel}) to set the values of different attributes as shown in Figure \ref{fig_system}. The details explaining the implications of different attributes are described in the following sections. The \textit{Graph View} window is used to display the social network as a node-link  diagram where the graph is laid out using a force directed algorithm proposed by Hachul and Junger \cite{hachul04}. Force directed algorithms are well suited for the visualization of community structures as the nodes densely connected to each other are pushed in close proximity and disconnected nodes are pushed far away. The \textit{Similarity Graph} window is used to display a graph of graphs, i.e. each node in this graph represents a graph where the x-axis represents the time line and the y-axis represents different values used to filter edges specified in the parameter \textit{Number of Slices}. This visualization is used to analyze the dynamics of the graph as it changes over time (details are explained in the following sections). The \textit{Cluster List} window contains a list of the clusters found in the graph. Clicking on one of the clusters in this list displays the contents of the cluster in the \textit{Cluster View} window. Clicking on a node in this graph displays the graph associated to this node in the \textit{Graph View} window. There is a small widget in the Similarity Graph window in the bottom left corner representing the minimum and maximum values of the strength metric and the color gradients associated to these values.

Finally the \textit{Hierarchy View} window is used to display the influence hierarchy extracted from the social network representing how influential a person is in the entire network. The layout algorithm used to display the hierarchy is called the Walker tree with improved implementation and was proposed by Buchheim \textit{et al.} \cite{buchheim02}. In addition to this hierarchical layout, the tree can be drawn using another layout known as the Radial Tree first introduced by \cite{kelly03}. The choice of the layout can be selected from the Attribute Panel where each layout has its own benefits. The Improved Walker layout helps to reveal the influence hierarchy as it is drawn top-down and the radial tree layout places the root at the center and the nodes connected to the root around it. Although this layout does not help to reveal the hierarchy but it does help to identify the central nodes as the leaves are placed far away from the center and the important nodes closer to the center in the layout.

The system is interactive where the size of each window can be changed. Zoom-in and zoom-out are associated with the scroll wheel of the mouse. The values in the attribute panel can be modified interactively where the corresponding graphs and their layouts change as the associated \textit{compute} or \textit{apply} buttons are clicked. If the value of the \textit{Clustering}($\tau$) is changed, the new clustering is calculated as soon as the \textit{scroll bar} is released.

The proposed framework consists of four major steps which are discussed in details below.

\subsection{Graph Discretization} \label{Discretization}

Once the data is loaded in the system, the first processing step of the proposed system is to convert the dynamic graph $G$ into a set of graphs representing snapshots of the graph at different time intervals. From $G$, we obtain a sequence of snapshots $G_{[0, \epsilon]},\ldots,G_{[T - \epsilon, T]}=G_1,\ldots,G_{\alpha}$, where $\alpha$ is the number of graphs obtained and $\epsilon$ is the discretization factor. The graph $G_{[t, t+\epsilon]}$ is the static snapshot corresponding to the time interval $[t, t+\epsilon]$ (i.e. the graph containing all vertices and edges involved during the time period $[t, t+\epsilon]$). The total number of graphs generated this way are equal to the total time period $[0, T]$ divided by the discretization factor $\epsilon$ which we represent by the factor $\alpha$. The system allows the user to set the value of $\epsilon$ which depends on the granularity of the time stamps present in the data set. 

%**Lets consider an example, if we consider the graph of scientists co-authoring an article having the year of publication as the temporal information, the user can choose a \textit{value} of 2 with the \textit{unit} set to year as shown in Figure\ref{fig_system}. So for a author graph, if we have the data from say year 2000 to 2009, as a result of the parameters set, we will get graphs for the year 2000-01, 2002-03 till 2008-09. Once these parameters are set, the user is required to click on the \textit{apply} button to activate the discretization process. As a result we obtain graphs for each time interval, we consider these graphs to be static as the edges remain constant during the entire interval of time.

Recall that the input data allows a relationship\_strength to be specified for each relationship. We calculate metrics using this value that can be associated to each edge. Currently the system provides three different calculations: the \textit{Total time}, \textit{Average time} and \textit{Occurrency}. \textit{Total time} refers to the commulative sum of relationship\_strength for all occurrences of nodes $(u,v)$. The \textit{average time} is the average calculated for all the instances and the \textit{Occurrency} is the frequency of occurrences of a relationship between any two nodes $(u,v)$. Use of these metrics depend on the data sets and the user's interpretation of values associated to a relationship. The user is required to select the type of metric from the \textit{Metric} drop down menu. 

%After selecting a metric, 
The system provides a method to filter edges having weak relationships. Since we cannot set a predefined threshold, we set multiple values to filter out edges. The user is required to input the \textit{Number of slices} ($\omega$) which is a positive integer. The system takes the minimum and the maximum values of the calculated metric and divides this range into slices as specified by the parameter $\omega$. For each graph in $G_{[0, \epsilon]},\ldots,G_{[T - \epsilon,T]}= G_1,\ldots,G_{\alpha}$, we obtain $\omega$ graphs. Finally we get $\omega \times \alpha$ graphs which are drawn in the bottom left window of the system as shown in Figure \ref{fig_system}. Each graph is represented as a node where the placement of the nodes on the x-axis represents the different time intervals ($\alpha$) and the y-axis represents the number of slices ($\omega$).  We call this graph the Similarity Graph as each node represents a graph and the nodes are placed in a grid layout. Clicking on a node displays the contents of the graph in the top left window as shown in Figure \ref{fig_system}. We explain how this layout helps in evaluating the similarity in the following sections.

%**As an example, consider the phone call data, lets suppose the minimum \textit{total time} a person called another person is 10 minutes and the maximum is 410 minutes. If we divide this into five slices specified in the parameter \textit{Number of slices}, we will have edges with five different ranges i.e. all the edges from total time 10 to 410 minutes. The second slice will contain only edges that have a total time between 90 and 410 minutes and so on where the final slice will contain edges where the total time is between 330 and 410 minutes. Once the slicing parameters are speicified, the user is required to click on the apply button.

%**So for example for a phone call data set, the total time that was spent during conversation between two nodes during a particular time interval is calculated and associated to each edge. In case of Co-authorship network, the number of articles co-authored in the time interval can also be used a the total time. 

%\input{Sections/decomposition}
\subsection{Graph decomposition}\label{decompo}

The input to the graph decomposition step is the set of graphs obtained as a result of the previous step. The basic idea is by considering two snapshot graphs corresponding to successive time intervals, they should have ``similar'' topologies if the dynamic graph does not undergo drastic changes between the two time intervals. To capture these topologies, our approach uses a decomposition algorithm by clustering the graph into smaller components. We describe the details of the decomposition process below.
 %and then comparing the clusterings of graphs of successive time intervals to detect major changes in the network over time.

\subsubsection{Strength metric}
Our decomposition algorithm is based on the \emph{Strength} metric, introduced by Auber \emph{et al.} \cite{ACJM03}. This metric quantifies the neighborhood's cohesion of a given edge and thus identifies if an edge is an intra-community or an inter-community edge. The \emph{strength} of an edge $e$ given by $w_s(e)$ is defined as follows:
%\begin{equation}
 $$ w_s(e) = \frac{\gamma_{3,4}(e)}{\gamma_{max}(e)}$$
%\end{equation}
where $\gamma_{3,4}(e)$ is the number of cycles of size $3$ or $4$ the edge $e$ belongs to, and $\gamma_{max}(e)$ is the maximum possible number of such cycles. Finally, one can define the \emph{strength} of a vertex as follows: $$w_s(u) = {\frac {\sum_{e \in adj(u)} w_s(e)}{deg(u)}}$$ where $adj(u)$ is the set of edges adjacent to $u$ and $deg(u)$ is the degree of vertex $u$. The time complexity to calculate the strength metric over all vertices (V) and edges (E) is $O(|E|\cdot(deg_{max})^2)$ where $deg_{max}$ is the maximum degree of the graph.
% (cf figure~\ref{fig_strength_exist}).
% \begin{figure}[ht]
% 	\centering
% 	\includegraphics[width= .4\linewidth]{FIGURES/decomposition/strength.eps}
% 	\caption{\label{fig_strength_exist} Strength metric of edge $e=(u,v)$ is computed by comparing the number of cycles of size $3$ and $4$ to the maximal number of such cycles..}
% \end{figure}

\subsubsection{Maximal independent set extraction}
%\begin{figure}[ht]
%	\centering
%	\includegraphics[width= .8\linewidth]{FIGURES/decomposition/misf.eps}
%	\caption{\label{illustr_misf}(a) Sub-graph of the "Hollywood graph" where vertices represent actors and two vertices are linked by an edge if the corresponding actors together in a movie. Vertices are colored according to their Strength value (from yellow for the lowest to dark blue for the highest). (b) and (c) are results obtained by our algorithm using two different approaches to extract communities centers.}
%\end{figure}

To identify the \emph{center} of communities within the network, we use a method inspired by \emph{MISF}~\footnote{Maximal Independent Set Filtering} \cite{GK00} where we extract a maximal set {\LARGE $\nu$} of vertices such that $\forall u,v \in $ {\LARGE $\nu$}, $dist_G(u,v) \geq 2$. The advantages of this algorithm are twofold: first, it gives the number of clusters with respect to the topology of the network and secondly, this technique guarantees the uniqueness of each found cluster (i.e. two clusters found by our approach cannot be identical) since a \emph{center} can only belong to one cluster.

Notice that since the vertices in {\LARGE $\nu$} are the \emph{center} of communities, these vertices should not be the pivots of the network as this may lead to over fitting a large community instead of several smaller communities. The network pivot nodes can be identified by low strength values as they are shared by several communities. Therefore, vertices with high strength values have to be added to the set {\LARGE $\nu$}. To extract such set, we use the algorithm~\ref{alg::misf}. 

 \begin{algorithm}[ht]
 \SetLine
 \KwIn{A graph $G=(V,E)$}
 \KwOut{A maximal set {\LARGE $\nu$} of vertices at distance at least $2$}
 vector$\langle$node$\rangle$ $sorted\_nodes$\;
 sortNodeWithStrength($G$, $sorted\_nodes$)\;
 \For{unsigned int $i$ from $0$ to (number of vertices in $G$)}{
 node $u$ = $sorted\_nodes[i]$\;
 \If{$u$ in $G$}{
 append({\LARGE $\nu$},$u$)\;
 \ForEach{node $v$ in neighborhood of $u$}{
 {remove($G$, $v$)\;}
 }
 remove($G$, $u$)\;
 }
 }
 \caption{Computation of the set {\LARGE $\nu$}. The sortNodeWithStrength ($G$, $sorted\_nodes$) method sorts the vertices by decreasing Strength values and store the result in $sorted\_nodes$.}
 \label{alg::misf}
 \end{algorithm}
 %In figure~\ref{illustr_misf}.(c), we used this approach to compute the centers of the communities resulting in three communities corresponding to the three movies of that network.
 
The time complexity of the sorting algorithm sortNodeWithStrength(Graph, vector$\langle$node$\rangle$) used within the algorithm~\ref{alg::misf}  is $O(|V| \cdot log(|V|))$. It is easy to show that the complexity of the \textbf{for} loop is $O(|V| + |E|)$. To compute {\LARGE $\nu$}, we sort (in descending order) the vertices $V$ according to their strength values as $V'$. Thereafter, we iterate over $V'$ adding the top node to {\LARGE $\nu$} and removing it and its neighbors from $V'$ until $|V'|$ = $0$. The complexity of this algorithm is $O(|V| \cdot log(|V|)+ |E|)$ in time and $O(|V|+ |E|)$ in space.

\subsubsection{Extracting communities}
We use the high strength node set {\LARGE $\nu$} to extract communities from the input network. The main idea is to build \emph{balls} with radius $1$ around the vertices in {\LARGE $\nu$}. For each node $u$ $\in$ {\LARGE $\nu$}, if an edge $(u,v)$ has a strength value higher than a given threshold $\tau$, then this edge is considered as an intra-cluster edge and the node $v$ is added to the community of $u$. The threshold $\tau$ is a function of the number of vertices and edges in the network. We consider several values for the threshold, $\tau_1,\dots,\tau_m$ obtaining $m$ different clusterings at each time interval. The time complexity of the communities extraction is $O(|E|)$ and its space complexity $O(|V| + |E|)$. The overall complexity of our decomposition algorithm is $O(|E| \cdot deg_{max}^2 + |V| \cdot log(|V|))$ in time and $O(|V|+|E|)$ in space. 

After the Graph decomposition step, we obtain clusterings for each graph in the Similarity graph. The user can set the value of $\tau$ from the interface using the slider where the range $[0,1]$ represents the metric strength calculated on edges as described previously. 

%A simple grid layout can be used to visualize $n \times m$ clusterings as shown in Figure~\ref{changes}. The time interval is represented on the x-axis and different values of $\tau$ on the y-axis.
%
%\begin{figure}
%	\centering
%	\includegraphics[width= .85\linewidth]{FIGURES/paolo_VASTcompressed.eps}
%	\caption{Similarity graph of the Catalano/Vidro network with maximum and average similarity at bottom. Major changes occurred between steps 2 and 3 and minor changes between steps 8 and 9}
%	\label{changes}
%\end{figure}

%\input{Sections/similarity_and_consensus}
\subsection{Detection of Changes}\label{detection}
We denote the clustering set by $C$ where each clustering $C_{i,j}$ corresponds to the decomposition of the graph $G_i$ with parameter $\tau_j$. As the decomposed graphs are naturally ordered with respect to time, the most probable cluster evolution can be found by comparing each $C_{i,j}$ with each $C_{i+1,k}$ $\forall i,j,k$ such that $1 \leq i < n$, $1 \leq j,k \leq m$. We describe a  \emph{similarity metric} in the next section to evaluate the similarity between each pair of clusterings in $C$.

\subsubsection{Similarity metric}
The \emph{similarity metric} aims to evaluate the similarity between two collections drawn over the same elements. It is related to the metric used in clustering protein-protein interaction networks \cite{Algorithm_evaluation}. The metric is based on the concept of \emph{representativeness}. We say that a cluster $c_a \in C_{i,j}$ is a good representative of a cluster $c_b \in C_{i+1,k}$ $iff$ $c_a$ contains a high ratio of the elements of $c_b$ and a small ratio of elements not in $c_b$. We define \emph{directed cluster representativeness} as:
\[ \rho_{c_a \rightarrow c_b} = c_a \cap c_b\ /\ |c_b| \qquad \rho_{c_b \rightarrow c_a} = c_a \cap c_b\ /\ |c_a| \]
which corresponds to the normalized ratio of the common elements between the two clusters.

We further define the \emph{undirected cluster representativeness}, or more simply \emph{cluster representativeness} as:
\[ \rho_{c_a,c_b} = \sqrt{ \rho_{c_a \rightarrow c_b} \cdot \rho_{c_b \rightarrow c_a} } \]
which corresponds to the geometrical mean of the direct representativeness of each cluster with respect to the other.

Next, we extend the definition of cluster representativeness to groups of clusters or clusterings. We say that $C_{i,j}$ is a good representative of $C_{i+1,k}$ if the former contains a good representative cluster for each cluster in the latter. As small size clusters tend to bias the representativeness values, we give more importance to clusters representative of larger size clusters over smaller ones. We define the \emph{directed clustering representativeness} as the weighted average (over the cardinality of the clustering) of the value of the best cluster representative found in $C_{i,j}$ for each cluster in $C_{i+1,k}$:
\[ \sigma_{C_{i,j} \rightarrow C_{i+1,k}} = \frac{\sum_{c_b \in C_{i+1,k}} \max_{c_a \in C_{i,j}} \rho_{c_a, c_b} \cdot |c_b|} {\sum_{c_b \in C_{i+1,k}} |c_b|} \]

Similarly, we define the \emph{undirected clustering representativeness} as the \emph{similarity metric}:
\[ \sigma_{C_{i,j},C_{i+1,k}} = \sqrt{ \sigma_{C_{i,j} \rightarrow C_{i+1,k}} \cdot \sigma_{C_{i+1,k} \rightarrow C_{i,j}} } \]

\subsubsection{Clustering Visualization}
Under the hypothesis that cluster evolution presents an inertia towards drastic changes (that means that clusters do not change drastically at each time step), the similarity between different clusterings helps to identify a better parameter value $\tau$. Currently, we are unable to select the optimal value $\tau_j$ that gives us the best clustering result for the graph $G_i$. Nevertheless, as a heuristic to estimate a good $\tau$, we detect a sequence of clusterings $C_{i,j},C_{i+1,k},C_{i+2,l} \dots$ that has a higher similarity metric at each step than the average.

To study the behavior at two successive time intervals, we calculate the maximum and the average similarity of two successive clusterings from $\sigma_{C_{i,j},C_{i+1,k}}$ $1 \leq j,k \leq m$. If the difference between the maximum similarity and the average similarity values is large, this signifies radical changes in the network whereas if the difference is small, we can infer that no significant changes occurred in the network between these two consecutive time intervals.
%Other conclusions can be deduced by the analysis of the similarity metrics for two following time steps. Looking at the values $\sigma_{C_{i,j},C_{i+1,k}}$, with $1 \leq j,k \leq m$, and calculating from them indicators like their maximum or their average, we can identify which of time steps $i$, $i+1$ present more or less variations. If we cannot find any good matching at all, then the graphs $G_i$, $G_{i+1}$ must be changed radically. On the other hand, if the mathings are almost all good, then there must have been really few changes between the two graphs.
To facilitate this analysis, we use a a visual representation of the similarity metric value computation between the evolving clusterings as shown in Figure~\ref{fig_system} shows the similarity metric computation and network changes for the Co-authorship data set. For each clustering $C_{i,j}$, we add an edge corresponding to clusterings compared through the similarity metric with $C_{i,j}$. These edges are then weighted with the similarity metric $\sigma$ and graphically displayed using a varying color scale and/or varying edge thickness. %Additionally, the maximum and the average values of the similarity metric at each time step is represented as a linear graph at the bottom as shown in figure~\ref{changes}.

To visualize each cluster, the user can select a node in the \textit{Similarity Graph} window, all the clusters present in this graph are listed in the \textit{Cluster List} window. We can explore these clusters individually by selecting a cluster in the list and visualize its contents in the top right window as shown in Figure \ref{fig_system}.

\subsubsection{Community Extraction}
With time, communities can expand to include new nodes or merge with other communities or decrease in size by deleting nodes or splitting into subgroups. Thus, a community at a given time step might appear as two distinct groups either due to a previous split or a pending merge. To overcome this problem and obtain a global idea of the community composition, we compute the consensus communities in the input network. At each time step, each community is represented by clusters detected. As these clusters represent a snapshot of the communities, we can follow the community evolution by matching the clusters between consecutive time intervals.

Let $\overline{C}_x,\overline{C}_{x+1},\overline{C}_{x+2}\dots$ be the clusterings $C_{x,j},C_{x+1,k},C_{x+2,l}\dots$ along a similarity path. We know the similarity metric $\sigma$ for each pair of consecutive $\overline{C}_i,\overline{C}_{i+1}$ clusterings and therefore the clustering representativeness between each $c_a \in \overline{C}_{i}$ and $c_b \in \overline{C}_{i+1}$. Thus, we use these values to match the clusters, and identify the clusters $c_b$ that are representative of clusters $c_a$.

Calculation of consensus communities between two graphs of different time intervals is achieved by first selecting a node in the similarity graph by clicking it, and then selecting another node by holding the \textit{ctrl} button and clicking the second node. The path between these two nodes with maximum similarity is calculated along with the corresponding consensus graph.
%Thereafter, the consensus communities can be detected from the collection of matched clusters generated above. We use different filtering algorithms---union, intersection, fraction threshold---the optimal choice depends on the context and the properties of the input data set.

The computation of $\sigma$ between two clusterings $C_{i,j}, C_{i+1,k}$ is bounded by the computation of the intersection between each pair of cluster $c_a \in C_{i,j}$, $c_b \in C_{i+1,k}$. This step requires at most $Q^2 |V|$ comparisons where $Q$ is the maximum cardinality of $C_{i,j}$ and $V$ is the number of nodes in the network.
%Let us assume we are analysing a network of $p$ nodes, over $n$ time steps each having one $m$ sets of parameters. We have in total $n \times m$ clusterings.
As each clustering $C_{i,j}$ is compared with all the clusterings at the following time steps, we calculate the similarity metric $(\alpha-1)\omega^2$ times. Typically $\alpha$ and $\omega$ are not very large---thus, the overall complexity is acceptable for an interactive computation. The calculation of the consensus communities depends on the filtering algorithm chosen by the user, but it is generally bounded by the computation of the similarity graph.

\subsection{Influence hierarchy}\label{hierarchy}

We define the \emph{influence hierarchy} as a tree $G_{T}$ = $(V_{T},E_{T})$ where $V_{T} \subset V$ is a subset of vertices in the social network. The height of a node $v$ $\in$ $V_T$ represents the strength of influence of that node in the network. Our technique is based on the Delta efficiency metric \cite{ML06,larsen07} which computes the importance of a vertex with respect to the flow of information in the entire network. To quantify the efficiency with which the nodes in the network exchange information, we use the idea of~\cite{LM04} to calculate this efficiency. We know that all nodes exchange information over a network represented by a graph $G=(V,E)$, and this information can be picked by other nodes if required. For the experimental data sets, each cell phone call or each co-authored artifact represents such an exchange of information. The communication efficiency of the network $\varepsilon_{ij}$ between the nodes $i$ and $j$ is inversely proportional to the shortest path in the graph between $i$ and $j$: $\forall i,j \in V$ $\varepsilon_{ij} = 1/d_{ij}$, where $d_{ij}$ is the shortest path between $i$ and $j$. If there is no path between $i$ and $j$, then $d_{ij}= + \infty$ and $\varepsilon_{ij}= 0$. 
We can quantify the efficiency of the whole network be calculating $\varepsilon_{ij}$ for each and every pair of nodes. The average efficiency of the graph $G$ can be defined as:
\begin{eqnarray*}
Eff(G)=\sum_{i \neq j \in V} \varepsilon_{ij}\ /\ |V| \cdot (|V|-1)
\end{eqnarray*}

This metric gives us the communication efficiency of the network. To find a hierarchy in the network, we need to evaluate the  efficiency or the \emph{criticality} of each node as proposed by~\cite{LM04}. The idea is that if an important member of the network is removed, the efficiency of the graph should decrease. We define the Delta Efficiency(DE) of a node as:
\begin{eqnarray*}
I(node_i) = \Delta Eff_i = Eff(G) - Eff(G\backslash \{i\}) 
\end{eqnarray*}

once we have the Delta Efficiency of each node, we can use this efficiency to assign weights to edges. There are several ways to assign a weight to an edge if the nodes are weighted. One way is to take the average of the two nodes connecting an edge. Currently we use the maximum of the two nodes connecting an edge given by the equation:
\begin{eqnarray*}
Weight(e_{ij})= Max\{\Delta Eff_i,\Delta Eff_j\} 
\end{eqnarray*}
The term $e_{ij}$ refers to an edge between nodes (i,j) and $\Delta Eff_i$ and $\Delta Eff_j$ are their respective delta efficiency values. From this weighted graph, we use Kruskal's Minimum Spanning Tree algorithm~\cite{K56} to generate a tree. This tree reveals the influence hierarchy in the network by selecting the node having the highest delta efficiency value. The tree can be calculated by clicking on the \textit{compute} button and visualized in the \textit{Hierarchy View} window.

From the given problem statement of the Catalano/Vidro data set \cite{bourqui09}, we were required to find people with specific roles in the social network. The \textit{Boss}, his \textit{right hands} and his \textit{brother}. As compared to generic social networks, in a terrorism network, the leader tries to hide himself in the network and does not have a high delta efficiency value. The leader is usually in contact with only a few people who are responsible of diffusing information in the entire network which we call \textit{right hands}. Calculating the hierarchy to find the leader in this network requires some adaptability which can be activated using the \textit{counter terrorism mode} from the attribute panel by clicking the \textit{check box}. The details of how the hierarchy is calculated in this mode are explained in the case study (see section \ref{case}).

\section{Case Study}\label{case}

\subsection{Co-authorship Network data set} 

\begin{figure}
	\centering
	\includegraphics[width= .59\linewidth]{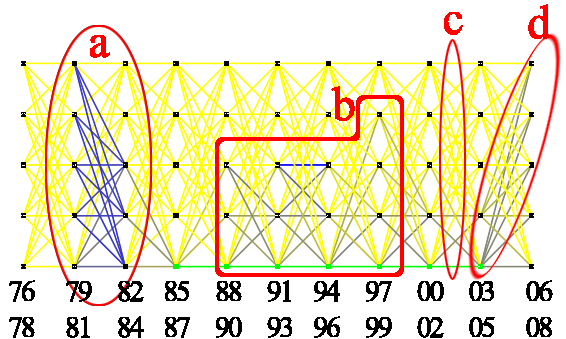}
	\caption{Similarity graph for Co-authorship network with focus on \textit{Ulrik Brandes} where blue color represents perfect similarity and yellow represents very low similarity. (a) suggests very high similarity between time intervals 1979-81 and 1982-84. (b) suggests a long period of high similarity between 1988 and 1999. (c) marks a major structural change as between this period, the similarity drops to a very low value. (d) resurgence of the similarity between the periods 2003-05 and 2006-08.}
	\label{similaritybrandes}
\end{figure}

As described previously, the co-authorship network was constructed by taking \textit{Ulrik Brandes} as the starting point and all the researchers connected to him at distance 2. The \textit{Similarity Graph} shown in Figure \ref{similaritybrandes} marks the important areas that can be used to deduce interesting information. The enclosed area labeled (a) refers to the similarity between two consecutive time intervals, 1979-81 and 1982-84 represented by blue colored edges between the graphs of this period. The second interesting period is labeled (b) where high similarity between 1988 and 1999 can be noticed. This is the period where we find the first publication of \textit{Ulrik Brandes}(1997). He finished his Ph.D. thesis in the year 1999. As the graph contains people closely connected to him, they are people working in the same domain (graph drawing and information visualization) and probably most of them are from the same university where \textit{Ulrik Brandes} did his doctorate, justifying the high similarity between graphs of this period. Research teams work on similar topics and there is always a tendency that while being at the same university, people tend to work with the same group of people with whom they have worked earlier. Clustering the graph for these periods would result in more or less the same clusters and stability over a time period in the collaboration behavior of researchers.

Between the year 1997-99 and 2000-02, there is a decrease in the similarity values which continues to decrease between the period 2000-02 and 2003-2005 as labeled (c) in the figure. During this period, \textit{Ulrik Brandes} first moved to the University of Sydney and then to Brown University for post-doctoral research fellowships. and thus the similarity values show a major structural change between these periods. From the year 2003, he has been working at the University of Konstanz and thus the similarity between the period 2003-05 and 2006-08 represent stability in the collaboration pattern of \textit{Ulrik Brandes}.

From Figure \ref{fig_system}, the hierarchy of this network is visible where we find \textit{Ulrik Brandes} at the root which is consistent with the way the data set was collected. Since he plays the role of the central person for information interchange between the rest, it is no surprise that we find him as the person with the highest delta efficiency value. The tree has a depth of 2, again an implication of the way the data set was collected since we consider only authors lying at distance 2 from \textit{Ulrik Brandes}. On the other hand, tree is very wide since he is an author who has collaborated with many people. These people themselves have collaborated with many people and such an example is depicted in Figure \ref{fig_system} where \textit{Michael Kaufmann} (additional reviewer of Ph.D. thesis of \textit{Ulrik Brandes}) is highlighted in the \textit{Graph View} and the \textit{Hierarchy View}. %From the \textit{Cluster List} view, one of the cluster is selected where the only node represent the stablity of \textit{Micheal Kaufmann} during the period  

\subsection{Catalano/Vidro set} 

To visualize the hierarchy of the Catalano/Vidro network, which is a terrorist network, the user is required to activate the \textit{counter terrorism mode}. The idea behind this mode is based on the study of social networks, we know that there are three kinds of roles in a network---the \emph{leaders}, who are the thinking heads; the \emph{gatekeepers}, who control the diffusion of information within the network and the \emph{followers} who just execute orders. The ones that have the largest activity within the network are the gatekeepers and therefore, they have the highest delta efficiency values. On the other side, leaders and followers have very restricted communications (leaders just issue orders while followers receive/execute orders) which is the reason why they have very low delta efficiency values. Past work has shown that leaders try to hide themselves among followers~\cite{ML06,larsen07} (due to low delta efficiency values for both) to escape detection. One of the primary goals in hierarchy detection is to distinguish between followers and leaders. Since there are three roles that constitute a hierarchy in the social network, we do this by finding a tree representing this hierarchy such that it reveals the leaders and the followers.

Our inspiration comes from the fact that the edges of this tree must be part of the input social network. We know that the gatekeepers (\textit{right hands}) of the leaders (\textit{boss}) are the most critical nodes in the hierarchy. By definition, they are very close to the leader nodes in the network. We infer the hierarchy by using the spanning tree of the modified network where each edge is weighted by the importance of the relationship between the two nodes using delta efficiency as the importance metric. Thereafter, we classify the different role types within the influence hierarchy computed from the spanning tree.

\begin{algorithm}[t]
{\small
\SetLine
\KwIn{A graph $G=(V,E)$, DE $n.\Delta$ of each vertex $n$, the spanning tree $T$}
\KwOut{The root node $boss$ of the hierarchy and an orientation of the spanning tree $T$}
int $number\_of\_nodes$ = $3$ / $100$ x $|V|$ \;
Sort($V$, rule $>$, $\Delta$) \;
\For{$i$ from $1$ to $number\_of\_nodes$}{
  table $neighborhood$ = $V[i]$.getNeighborhood \;
  \For{$j$ from $1$ to $neighborhood$.size}{
    $M[i][j]$ = $neighborhood[j]$\;
}
}
list $result$\;
\For{$i$ from $1$ to $number\_of\_nodes-1$}{
  \For{$j$ from $(i+1)$ to $number\_of\_nodes$}{
    \For{$k$ from $1$ to $M[i]$.size}{
      \For{$l$ from $1$ to $M[j]$.size}{
        if( M[\(i\)][\(k\)] == M[\(j\)][\(l\)]) then result.push(M[\(i\)][\(k\)]) ;
}
}
}
}
sort($result$, rule $>$, $nodes\_id$) \;
node boss = maxTimeAppears($result$) \;
makeOrientedTree($T$, $boss$) \;
\BlankLine
}
\caption{Inferring the network hierarchy.}
\label{alg::hierarchy_two}
\end{algorithm}

We associate a weight with each edge between two nodes that is the difference between the delta efficiency value of the connected nodes. A high difference between delta efficiency values indicate that the two connecting nodes should not be placed on the same level in the hierarchy so the edge between these nodes can be removed. We take the absolute value that is inversely proportional to value associated to the edges as the edge weights for Kruskal's Minimum Spanning Tree algorithm to compute the hierarchy tree. The complete algorithm is listed as \ref{alg::hierarchy_two}.

\begin{figure}
	\centering
	\includegraphics[width= .59\linewidth]{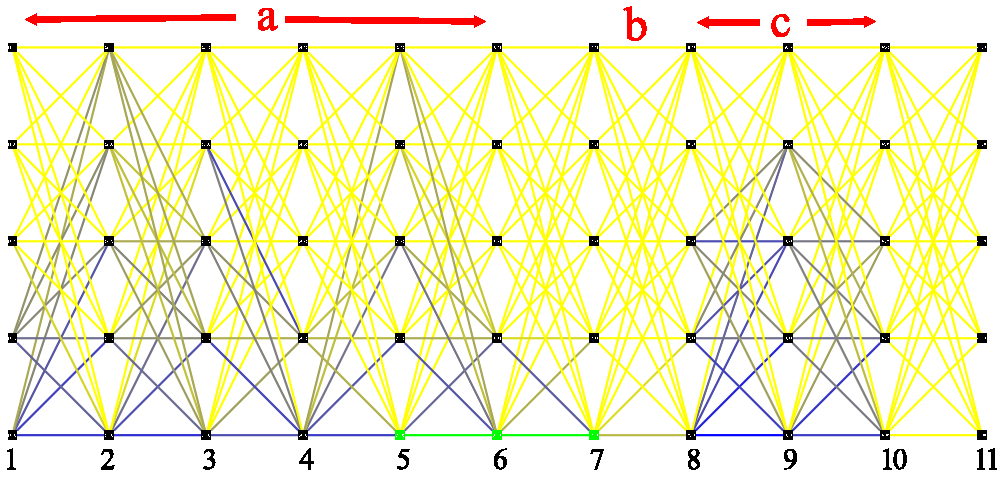}
	\caption{Similarity graph for Catalano/Vidro network over a 10 day period where blue color represents perfect similarity and yellow represents very low similarity. (a) suggests high similarity between day 1 to day 6. (b) suggests a high dissimilarity period between day 7 and 8 marking a major structural change as between this period, the similarity drops to a very low value. (c) the similarity between day 8 and 10 is high again representing the stability of the structure after the change.}
	\label{similarityvast}
\end{figure}

From Figure \ref{similarityvast}, we can easily find that a major structural change occurred during day 7 and day 8. The terrorist network actually changes their mobile phones destroying the old communities and forming new ones during day 8 and day 10. Thus the visualizing the change in the cluster similarity helped us to identify an interesting information.

\begin{figure}
	\centering
	\includegraphics[width= .35\linewidth]{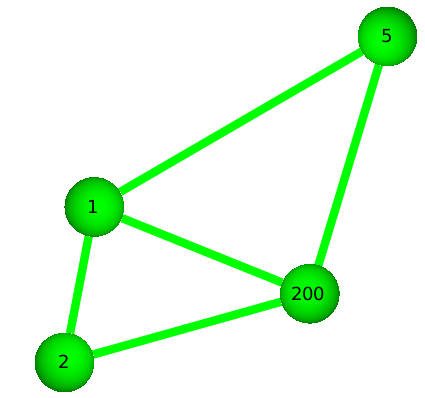}
	\caption{Catalano/Vidro Data Set. A cluster found in the period day 1 to day 6 where all the high efficiency nodes labeled 1, 2, 5 communicated with a node 200. The high efficiency nodes represent the \textit{right hands} of the leader and the node 200 is in fact the leader of the terrorist network. Since the communication between the leader and the right hands over a period of time remains stable, thus the system identified these nodes as a single cluster where as the people who play the role of \textit{followers} keep changing as a function of criminal activity.}
	\label{vastleader}
\end{figure}

Figure \ref{vastleader} represents one of the clusters found in during the period day 1 and day 6. The highest delta efficiency nodes in this network were identified to be the nodes labeled 1, 2 and 5. Since we expected that the \textit{right hands} of the leader (\textit{Boss}) are responsible in diffusing information in the terrorist network, they will have the highest delta efficiency values. This information was proven to be correct as these nodes indeed represent the \textit{right hands} of the \textit{Boss}. Moreover, we suspected that all the right hands communicate with the \textit{Boss}, and over a long period suggesting that there is no change in either the \textit{Boss} or his \textit{right hands}. The only cluster having the right hands and a single node communicating to all of them was the cluster shown in Figure \ref{vastleader}, thus revealing the identity of the \textit{Boss} which is the node 200.

%\subsubsection{Inferring the hierarchy}

Once we have the spanning tree, the next step is to find the leaders (\textit{boss}) in the network. We know that the \textit{boss} has a low delta efficiency value as he tries to hide himself by communicating less in the network and that he is in the neighborhood of the gatekeepers (\textit{right hands}). \cite{LM04} suggests that these \textit{right hands} have the highest delta efficiency values in the network. But it may not be true that all the highest valued nodes are the right hands of the boss. To find the correct number of right hands of the boss we use a heuristic value of $3\%$ that worked well for the input data set. The idea behind finding the \textit{boss} is that we take the $3\%$ nodes having the highest delta efficiency value and construct a separate neighborhood list for each of these nodes which contains their immediate neighbors. Once we have these lists, we count the number of times the elements that appear in both the lists by taking two lists at a time. For example consider two node lists $list\ 1$ and $list\ 2$ that have $3$ elements that are present in both of these lists. We add a value $1$ as count to each of these elements. We repeat this process for all the possible combinations of the lists and at the end we come up with the node that is being communicated the most by the right hands. This node is probably the \textit{boss} of the network as we know that the \textit{boss} communicates a lot with his \textit{right hands}. In case if two or more nodes have the same count, we take the node with the lower delta efficiency value as we know that the \textit{boss} does not have a high delta efficiency value.

Once we have a spanning hierarchical tree and the \textit{boss} of the hierarchy, we adjust the orientation of the tree, starting from the root (boss) to have an ordered hierarchy. Given the complexity of our naive implementation of delta efficiency $O((n+1)(m+n^2))$ and the complexity of algorithm \ref{alg::hierarchy_two} is $O(n^4)$, the overall worst-case complexity is $O(n^4)$ where $n$ is the number of nodes and $m$ is the number of edges in the consensus graph. In practice, however, the algorithm performs better than the worst-case. 

Once the hierarchy is generated, it is displayed in the \textit{Hierarchy View} window as shown in Figure \ref{fig_system}. As described previously, the nodes with high influence in the network are placed higher in the tree and the lowest influence nodes being at the leaves.

\section{Conclusions and Future Research Directions}\label{conclusion}
In this paper, we have presented a system to analyze dynamic social networks to detect changes over a time period based on graph discretization and clustering. We have also presented a method to discover influence hierarchy in social networks using communication efficiency and minimum spanning tree algorithm. We have applied our system on two different data sets and obtained satisfactory results as we were able to correctly identify the time frames where major structural changes occurred as well as discover the influence hierarchy of the important people in the social networks. The system has some obvious limitations. As the number of graphs generated in the discretization step increases, the systems performance highly depends on the clustering algorithm used to cluster these individual graphs. If less graphs are generated, information loss can occur, thus we have a fine 

There are several details that we would like to address as part of future research to improve the overall system. Currently the discretization step divides the time interval into discrete windows of time which do not overlap. In case if a structural change occurs in the middle of a time interval, we are able to detect only the time interval and not the exact time instance. It would be interesting to analyze the data set if overlapping is allowed to facilitate the analysis of exact time instances where an event occurred. To cluster the graph, we have used the strength metric which is based on the topological information of a network. In the presence of a large number of attributes, we would like to incorporate other metrics that take into account these attributes which will certainly improve the quality of clustering. We use a hierarchical layout to display the influence hierarchy which can very well be replaced by other layouts such as ego-centric layout which will help us to focus on individuals and their roles in the entire network. All these questions present us with new challenges to analyze and understand the evolving social networks.

\bibliographystyle{abbrv}    

\bibliography{references}   % name your BibTeX data base

%\begin{thebibliography}{}
%%
%% and use \bibitem to create references. Consult the Instructions
%% for authors for reference list style.
%\end{thebibliography}

\end{document}